\documentclass[aps,prl,twocolumn,superscriptaddress,noshowpacs,longbibliography]{revtex4-2}
\usepackage{amsmath}
\usepackage{amssymb}
\usepackage{natbib}
\usepackage{bm}
\usepackage{times}
\usepackage{color}
\usepackage{graphicx}
\newcommand{\kB}{k_{\rm B}}
\newcommand{\xb}{{\bf x}}
\newcommand{\XX}{{\bf X}}

\newcommand{\dee}{{\rm d}}
\newcommand{\etab}{{{\bm \eta}}}

\newcommand{\calP}{{\mathcal P}}
\newcommand{\eqa}{\begin{eqnarray}}
\newcommand{\eeqa}{\end{eqnarray}}
\newcommand{\eq}{\begin{equation}}
\newcommand{\eeq}{\end{equation}}
\newcommand{\ex}{{\rm e}}

\begin{document}

\title{Stochastic transitions: Paths over higher energy barriers can dominate in the early stages}

\author{S.~P. Fitzgerald}
\email{s.p.fitzgerald@leeds.ac.uk}
\affiliation{Department of Applied Mathematics, University of Leeds, Leeds LS2 9JT, United Kingdom}
\author{A. Bailey Hass}
\affiliation{Department of Applied Mathematics, University of Leeds, Leeds LS2 9JT, United Kingdom}
\author{G. D{\'\i}az Leines}
\affiliation{Yusuf Hamied Department of Chemistry, University of Cambridge, Lensfield Road, Cambridge CB2 1EW, United Kingdom}
\author{A.~J. Archer}
\email{a.j.archer@lboro.ac.uk}
\affiliation{Department of Mathematical Sciences and Interdisciplinary Centre for Mathematical Modelling,
Loughborough University, Loughborough, Leicestershire LE11 3TU, United Kingdom}

\begin{abstract}

\noindent The time evolution of many physical, chemical, and biological systems can be modelled by stochastic transitions between the minima of the potential energy surface describing the system of interest. We show that in cases where there are two (or more) possible pathways that the system can take, the time available for the transition to occur is crucially important. The well-known results of reaction rate theory for determining the rates of the transitions apply in the long-time limit. However, at short times, the system can instead choose to pass over higher energy barriers with much higher probability, as long as the distance to travel in phase space is shorter. We construct two simple models to illustrate this general phenomenon. We also present an extension of the gMAM algorithm of Vanden-Eijnden and Heymann [J. Chem. Phys. {\bf 128}, 061103 (2008)] to determine the most likely path at both short and long times.

\end{abstract}

\maketitle

The transition dynamics of complex systems having many degrees of freedom can often be reduced to one or two reaction coordinates. These are the system degrees of freedom that evolve the slowest in time \cite{peters2017reaction}.
All the other (maybe very many) degrees of freedom are slaved to these slowest processes \cite{hanggi1990reaction}. The slow evolution of these systems is usually characterized by rare transitions between metastable states separated by significant energy barriers.
The identification of the reaction coordinates in high-dimensional (complex) systems remains extremely challenging \cite{rogal2021reaction}.
For example, for large molecules, the centre of mass is often a `slow' degree of freedom, whilst the fluctuations of the individual atoms within the molecule are the slaved `fast' degrees of freedom.
This is the case e.g.\ in biomolecular conformation changes such as protein folding \cite{cho2006p, dill2012protein}, nucleation-driven phase transformations \cite{blow2021seven}, chemical reactions in general and surfactant molecules in a liquid transitioning from being freely dispersed in the liquid or joined together in a micelle or adsorbing to interfaces \cite{israelachvili2011intermolecular, leal2007advanced, thiele2016gradient}.
An example of recent work to identify the relevant reaction coordinates is Ref.~\cite{appeldorn2022employing}, which uses machine-learning.
Of course, for high-dimensional systems, the energy landscape is often complex, with multiple critical points, barriers of various sizes and multiple transition paths connecting the stable states. For such systems, algorithms based on simplifying assumptions such as no barrier recrossings, single transition states, a smooth landscape, `long enough' (infinite) times often fail to provide an straightforward and accurate estimation of the rate \cite{blow2021seven} or to even identify the most likely path \cite{leines2016comparison, liu2022stability} under perturbations of the energy landscape.

We consider here a class of such stochastic dynamical systems where there is a simple choice of two transition pathways away from the initial state: one is over a smaller energy barrier (the activation energy barrier for chemical reactions), but the system has to evolve a greater distance in phase space (i.e.\ has a longer reaction pathway), while the other path is over a much higher energy barrier, but has a much shorter distance to travel in phase space.
Examples of such systems include where a surfactant molecule in liquid has a choice between adsorbing to an interface or forming micelles, or where a chemical reaction can proceed via a catalyzed or non-catalyzed route.
Standard reaction-rate theory (RRT) which includes transition state theory and other related approaches \cite{hanggi1990reaction} predicts that the path over the lowest barrier is the most likely and therefore dominates the dynamics, at least for simple energy landscapes.
However, even for these simple cases we find the standard RRT picture does not hold and the behaviour crucially depends on the timescale over which the system is sampled.
In particular, for shorter times (but still much longer than the timescale of the fluctuating `fast' degrees of freedom) the flux over the higher barrier can completely dominate the dynamics of the system and even at intermediate times, the transition probabilities are very different from the predictions of RRT approaches which do not consider the time taken; i.e.\ RRT only applies in the long-time limit.
The key finding of our work is that the length of time over which barrier crossing problems are allowed to proceed is critically important. In any system where the reaction is stopped after a certain time, the reaction pathway predicted by RRT may not be the one actually taken.
For example, this may be the case in flow reactors such as catalytic converters. However, in any system that can explore the long-time limit, the predictions of RRT are fully recovered.
Conversely, if the potential landscape and reaction coordinates are unknown, and are inferred via densities and rates measured from experiments or simulations necessarily performed on a finite timescale, then the dominant long-time dynamics of the system may be missed entirely.

Additionally, we develop a method for calculating the most likely path (MLP) through the potential energy landscape, useful for analysing systems with two or more dimensions.
Various techniques to compute the minimum energy (and hence most probable) path between two minima exist, including the string method and the geometric minimum action method (gMAM \cite{vanden2008geometric,heymann2008geometric}; see also \cite{koehl,olender}).
Such paths, sometimes known as the \emph{instanton}, are everywhere parallel to the potential gradient, and correspond to the infinite-time transition. In this work, we extend the gMAM approach to finite-time transitions, and derive a modified algorithm to compute finite-time, out-of-equilibrium paths.
These are no longer parallel to the potential gradient, and correspond to the most probable path conditioned on a finite duration.
We find that these can be radically different from the instanton, and may pass through completely different intermediate states.
We explain how these paths are connected to the full transient dynamics of the system given by the Fokker-Planck equation.

We demonstrate our findings with two simple generic toy models. The first is one-dimensional (1D) and the potential energy landscape has three minima. The system is initiated in the middle one and then has a choice to evolve either to the left or to the right. Our second model potential is two-dimensional (2D).
It has two minima and two saddles, meaning two different classes of path linking one minimum to the other.
One path is shorter, but over a high barrier in the potential, while the other is further, but over a much lower barrier.
RRT would suggest that the second is the dominant transition pathway, but we find that this is not the case if one only considers the system for sufficiently short times.
These systems are described by the overdamped stochastic equation of motion
\begin{equation}
\Gamma^{-1}\frac{d\xb}{dt}=-\nabla\phi(\xb)+\etab,
\label{eq:EOM}
\end{equation}
where $\xb$ is the `slow' relevant degree of freedom of the system (1D or 2D in the cases considered here), $\phi(\xb)$ is the potential energy of the system (strictly speaking in systems where irrelevant `fast' degrees of freedom have been integrated out, $\phi$ is a constrained free energy), $\Gamma^{-1}$ is a friction constant that henceforth we set equal to one (i.e.\ absorb it into the timescale) and $\etab$ 
is a random force originating from thermal fluctuations in the system. This is modelled as a white noise with zero mean $\langle\eta_i(t)\rangle=0$ and correlator $\langle\eta_i(t)\eta_j(t')\rangle=2 k_BT\delta_{ij}\delta(t-t')$, where $k_B$ is Boltzmann's constant and $T$ is the temperature (i.e.\ the amplitude of the random fluctuations).

The Fokker-Plank equation for the probability density $\rho(\xb,t)$ corresponding to Eq.~\eqref{eq:EOM} is \cite{gardiner1985handbook}
\begin{equation}
\frac{\partial\rho}{\partial t}=\Gamma\nabla\cdot\left[k_BT\nabla\rho+\rho\nabla\phi\right].
\label{eq:FP}
\end{equation}
When $\phi=0$ this becomes $\frac{\partial\rho}{\partial t}=D\nabla^2\rho$, the diffusion equation, with diffusion coefficient $D=\Gamma k_BT$. Note that Eq.~\eqref{eq:FP} can be written as a gradient dynamics
\begin{equation}
\frac{\partial\rho}{\partial t}=\nabla\cdot\left[\Gamma\rho\nabla\frac{\delta F}{\delta \rho}\right],
\label{eq:DDFT}
\end{equation}
with the Helmholtz free energy functional
\begin{equation}
F[\rho]=\int  \rho\left[k_BT\ln\rho+\phi\right]\dee^n\xb,
\label{eq:F}
\end{equation}
which is a Lyapounov functional for the dynamics. Note that these are the equations of dynamical density functional theory \cite{te2020classical, archer2004dynamical, marconi1999dynamic}.
For a given potential $\phi(\xb)$, the equilibrium density is $\rho(\xb)=\rho_0{\rm e}^{-\beta\phi(\xb)}$, where $\beta=(\kB T)^{-1}$ and $\rho_0$ is a constant determined by the normalisation of $\rho(\xb)$; i.e.\ $\rho_0^{-1}=\int {\rm e}^{-\beta\phi(\xb)}\dee^n\xb$.

When $\phi(\xb)$ has at least two minima, the quantity of interest is the typical waiting time to observe transitions between the minima.
Standard RRT states that this transition rate $k$ is given by the Arrhenius (or Kramers) relation $k=\nu\exp(-\beta\Delta\phi)$ where $\Delta\phi\equiv\phi(\xb_{s})-\phi(\xb_A)$ is the height of the barrier, with $\xb_A$ being the position of the minimum and $\xb_s$ the maximum (more generally saddle-point) on the barrier.
The prefactor $\nu$ depends on various factors \cite{hanggi1990reaction}, but it is the exponential that crucially determines the rate and can be thought of as originating from the ratio $\rho(\xb_s)/\rho(\xb_A)$, which is the probability of finding the system on the barrier divided by the probability of it being at the minimum.
However, this ratio $\propto\exp(-\beta\Delta\phi)$ only in the long time $t\to\infty$ limit. Solving Eq.~\eqref{eq:FP} with the initial condition $\rho(\xb,t=0)=\delta(\xb-\xb_A)$, we find that the RRT result can be completely wrong in some cases, if considering transitions with only a short time to occur.

\begin{figure}[t!] 
   \centering
   \includegraphics[width=0.9\columnwidth]{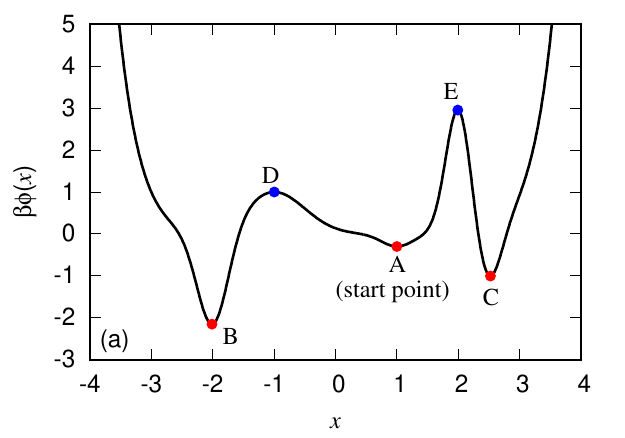}
   \includegraphics[width=0.9\columnwidth]{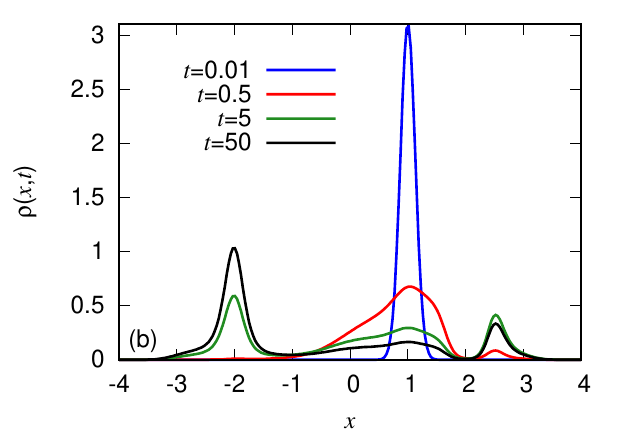}
   \includegraphics[width=0.9\columnwidth]{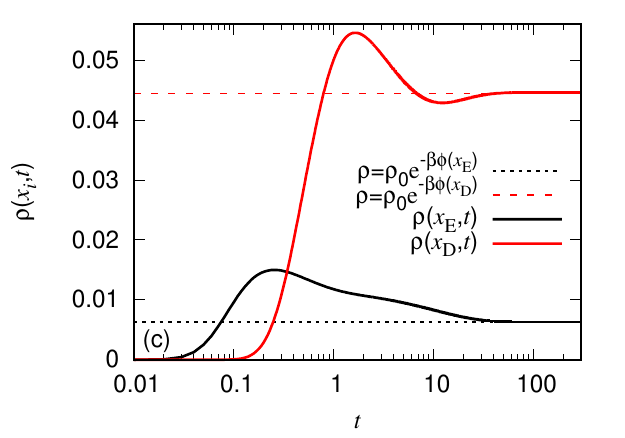}
   \caption{(a) The 1D model potential; (b) Probability density $\rho(x,t)$ over time as the system evolves in the potential, having started at point A at time $t=0$; (c) Density at the two saddle points versus time with equilibrium $t\to\infty$ values also shown.}
   \label{fig:1D}
\end{figure}

We consider first the 1D potential $\phi(x)$ in Fig.~\ref{fig:1D}(a); the equation for $\phi(x)$ is given in the supplementary information (SI).
This potential has 3 minima [labelled A, B and C in Fig.~\ref{fig:1D}(a)], at $x_B\approx-2$, $x_A\approx1$ and $x_C\approx2.5$ and two maxima [labelled D and E] at $x_D\approx-1$ and $x_E\approx2$.
We initiate the system in the minimum at A.
It can then either move to the right, over the much higher energy barrier at E, or it can go to the left over the lower barrier at D.
Going left, it has further to travel.

In Fig.~\ref{fig:1D}(b) we plot the density profile $\rho(x,t)$ obtained from solving Eq.~\eqref{eq:FP} for a sequence of different times $t$. Rather than initiating the system with the Dirac $\delta$-distribution centred at $x_A$, we use a narrow Gaussian corresponding to a free diffusion for the short initial time $t=0.01$.
By the time $t=0.5$ we see a sizable peak in $\rho(x,t)$ at C, the right hand minimum in $\phi(x)$, but very little density has made it to the minimum at B.
This is because B is further away, so in the early stages the system is more likely to cross the barrier at E, despite it being higher than the barrier at D.
It takes until $t\approx30$ for $\rho(x,t)$ to cease evolving in time and the system to reach the equilibrium distribution. Note also that at $t=5$ the density at C is {\it higher} than its eventual equilibrium value. Once the system has `found' the lower-energy minimum at B, density moves back over the high barrier at E to approach $\rho_0\ex^{-\beta\phi(\xb)}$.

In Fig.~\ref{fig:1D}(c) we plot the densities at the points D and E over time.
These are the locations of the two potential maxima (the barriers).
We see that at early times $t\sim0.1$ the probability of being at the highest maximum E is sizeable and well above the RRT probability $\sim \exp [-\beta\phi(x_E)]$, whilst the probability of being at the lower maximum D is still $\approx 0$, in contrast to the RRT prediction that the probability $\sim \exp [-\beta\phi(x_D)]$. Even at $t\sim 1$, the RRT predictions are still incorrect.

\begin{figure}[t!]
   \centering
 \includegraphics[width=0.9\columnwidth]{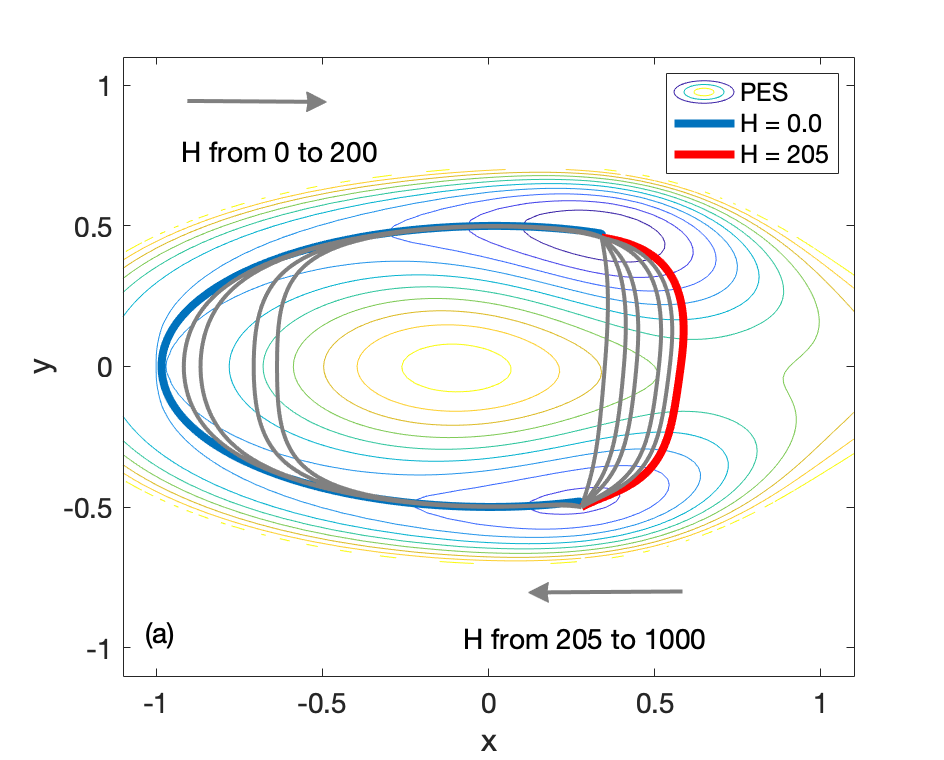}
 \includegraphics[width=0.9\columnwidth]{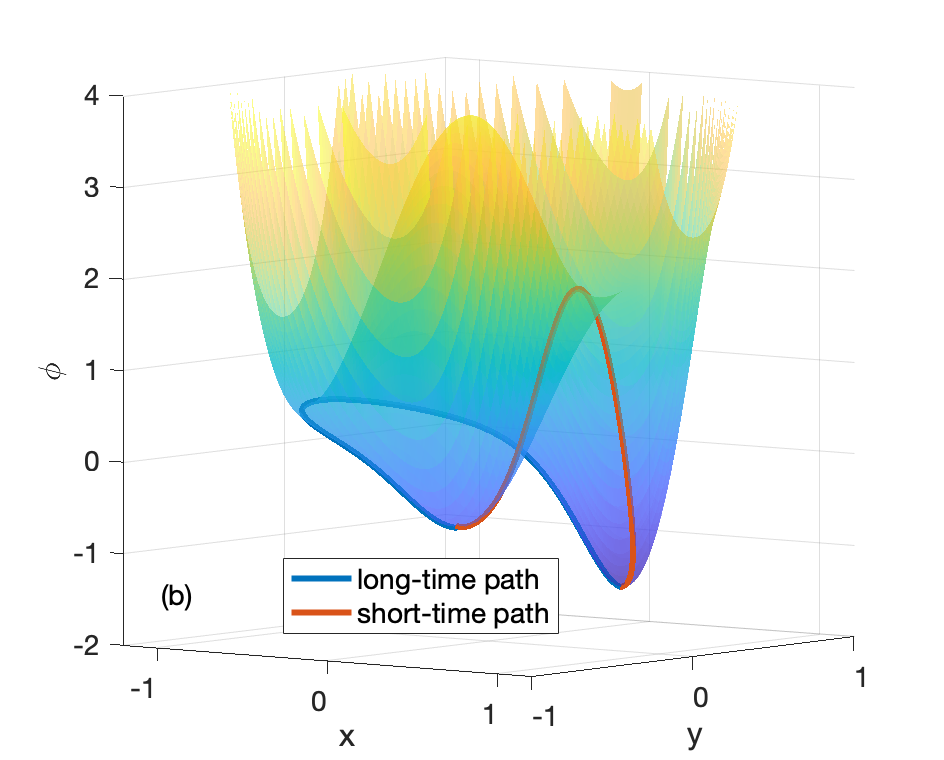}
 
 \vspace{3mm}
 \includegraphics[width=0.85\columnwidth]{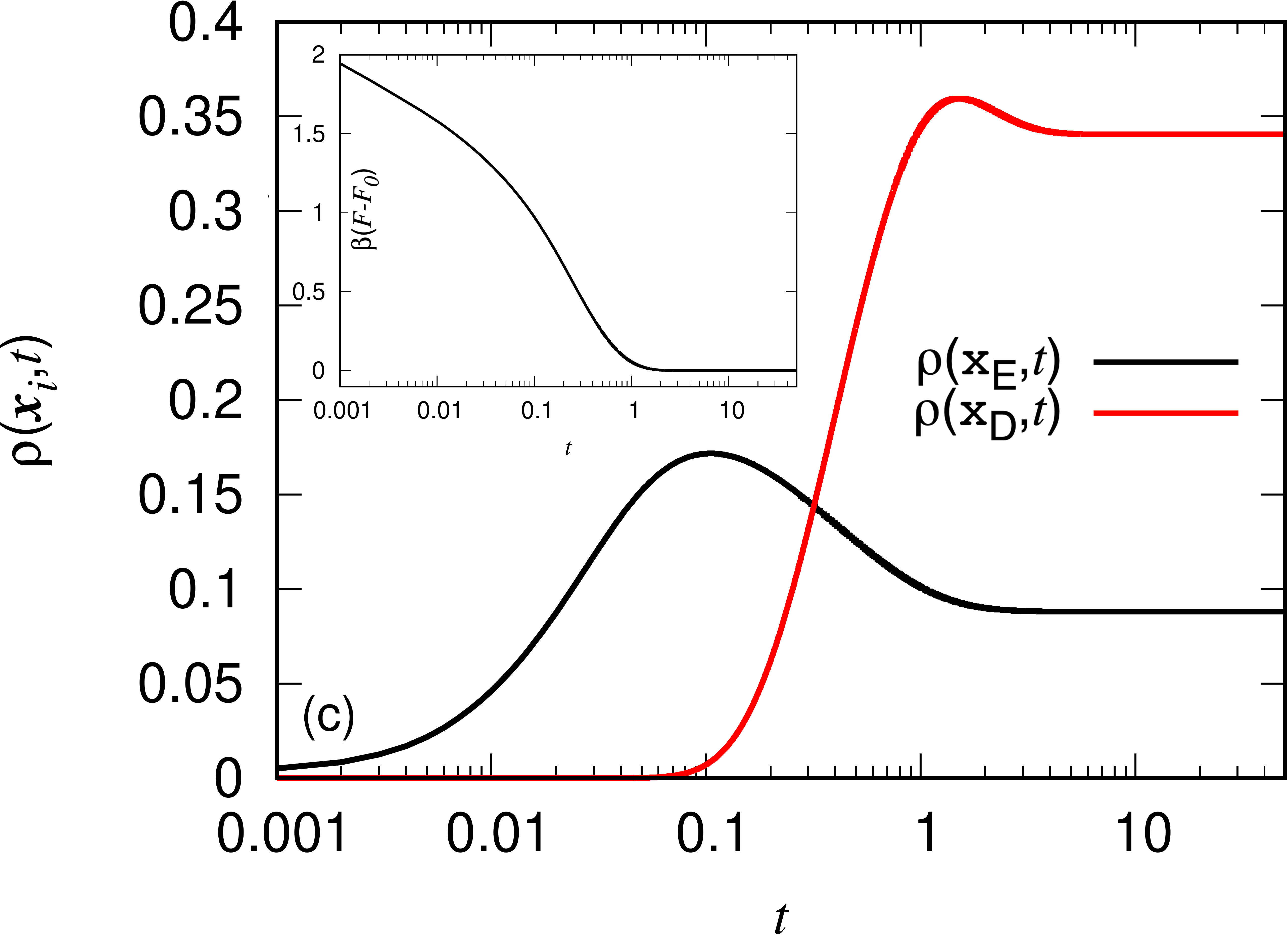}
   \caption{Panel (a) is a contour plot of our 2D potential $\phi$ together with modified gMAM results for the MLP, for various values of the path power $H$. Panel (b) shows a surface plot of $\phi$ together with the $t\to\infty$ long-time MLP (over the lower barrier) and the short-time MLP for $H=250$ (over the higher barrier). Note that the surface color scheme in (b) is the same as used for the contours in (a). In (c) we plot the density at the two saddle points, $\rho(\xb_D,t)$ and $\rho(\xb_E,t)$, as a function of time $t$. The inset of (c) is a plot of the Helmholtz free energy difference $(F-F_0)$ over time, where $F_0=F(t\to\infty)$.}
   \label{fig:2D}
\end{figure}

\begin{figure}[t!]
   \centering
 \includegraphics[width=1.0\columnwidth]{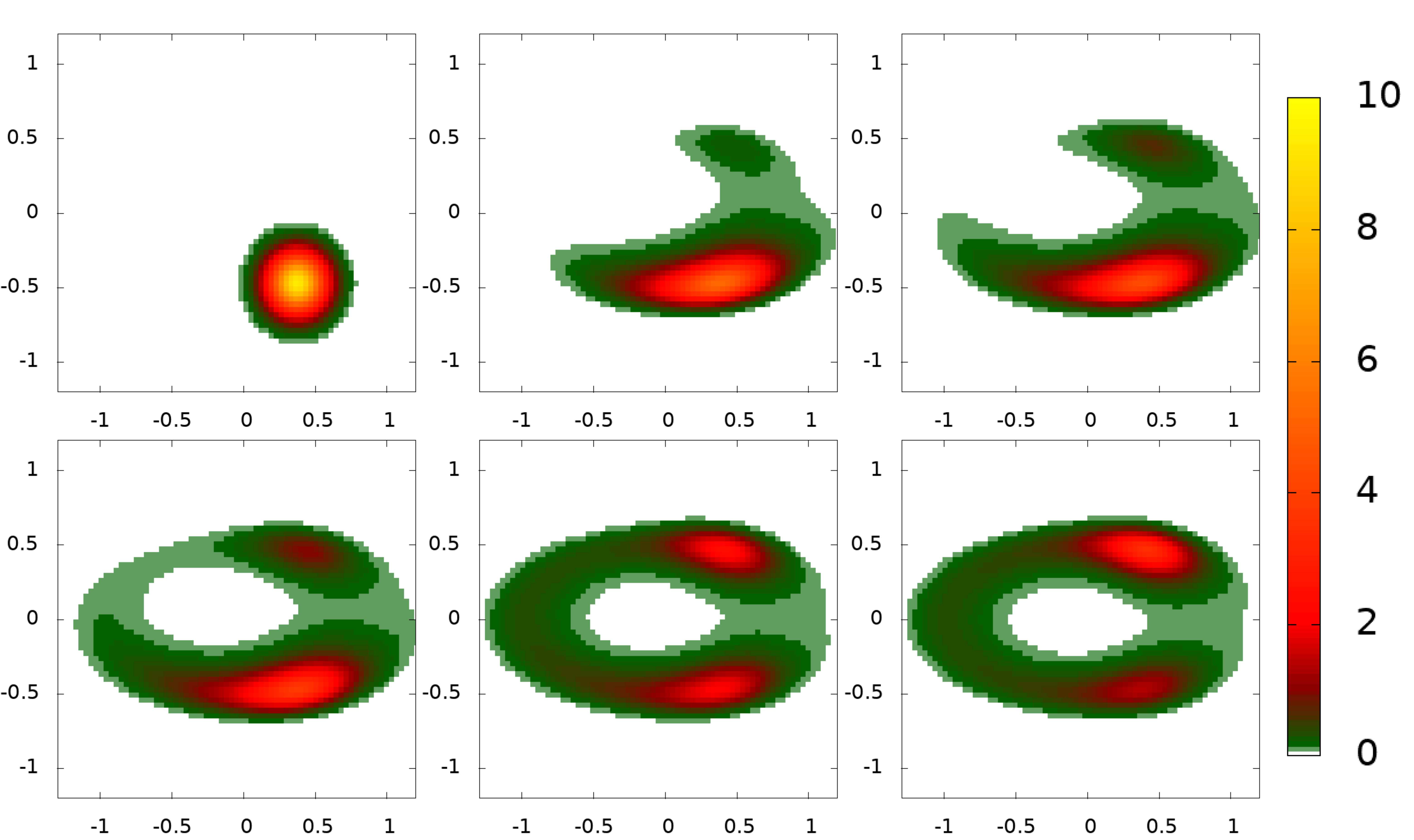}
 
   \caption{Density profile $\rho(x,y,t)$ at the times $t=0.01$, 0.02, 0.03, 0.04, 0.1, 5, going from top left to bottom right, for our 2D potential.}
   \label{fig:rhos_2D}
\end{figure}

We also consider a system evolving in the 2D potential displayed in Fig.~\ref{fig:2D}; the precise expression for this potential is given in the SI. Fig.~\ref{fig:2D}(a) shows a contour plot, whilst \ref{fig:2D}(b) is a surface plot. This potential has a local minimum at point A: $(x_A,y_A)=(0.38,-0.47)$ and the global minimum at B: $(x_B,y_B)=(0.42,0.47)$.
There is a local maximum near the origin. We initiate the system at A. There are two routes to go from A to B: the long route, round to the left in Fig.~\ref{fig:2D}(a), through the saddle at point D, $(x_D,y_D)=(-1,0)$, or the short route to the right through the saddle at point E at $(x_E,y_E)=(0.67,0)$.
The barrier height to the left is much lower, with $\beta\Delta \phi=1.4$. In contrast the barrier to the right is twice as high, with $\beta\Delta \phi=2.8$. Nonetheless, we see from Fig.~\ref{fig:2D}(c), which shows the density at the two saddle points (i.e.\ the tops of the two transition barriers) over time, that at early times the system is more likely to take the shorter route to the right, even though it is over the higher barrier. Note how similar Fig.~\ref{fig:2D}(c) is to Fig.~\ref{fig:1D}(c).
In the inset of Fig.~\ref{fig:2D}(c) we plot the free energy [Eq.~\eqref{eq:F}] over time, which as expected decreases monotonically over time.
In Fig.~\ref{fig:rhos_2D} we display the density profiles over time. These again show that at early times, the probability for the particle to be at the higher barrier at point E is much greater than at the lower barrier at D.

To elucidate this further we now generalize the gMAM algorithm to determine the MLP conditioned on finite transition times.
The Gaussian white noise $\etab$ in Eq.~\eqref{eq:EOM} has probability density functional 
$\calP[\etab]\sim\exp\left[-\frac{1}{4k_BT}\int_0^t\etab\cdot\etab \,\dee \tau\right]$. Substituting (\ref{eq:EOM}) into this, we immediately obtain
\begin{equation}
\calP[\xb]\sim\exp\left[-\frac{1}{4k_BT}\int_0^t|\dot{\xb}+\nabla \phi|^2 \dee \tau\right] \equiv \exp\left[-\frac{{\cal S}[\xb]}{4k_BT}\right]
\end{equation} for the probability weight attached to a path $\xb(\tau)$, and we have defined the {\it path action} $\cal S$. The transition probability $P(\xb_1,t|\xb_0,0)$ can now be written as a path integral \cite{wio,graham}
\begin{equation}
P(\xb_1,t|\xb_0,0) = \int{\cal D}\xb\, {\cal J}[\xb] \,\exp \left[-\frac{{\cal S}[\xb]}{4k_BT}\right],
\end{equation} where ${\cal J} = |\delta\etab/\delta \xb|$ is the functional Jacobian arising from the change of variables $\etab\to \xb$, and the integral is taken over all paths $\xb(\tau)$ with endpoints $\xb(0)=\xb_0$ and $\xb(t)=\xb_1$.
Note some constants are absorbed into the functional measure. This expression solves the Fokker-Planck equation \eqref{eq:FP} with initial condition $\rho(\xb,0)=\delta(\xb-\xb_0)$, but cannot be evaluated exactly except for simple special cases, such as quadratic $\phi$.
However, for paths realizing the transitions of interest, the action is typically much larger than $k_BT$, and so the integral is dominated by paths that minimize $\cal S$, i.e.\ paths that satisfy the Euler-Lagrange equations for $\cal S$:
\begin{equation} 
\ddot \xb = \nabla \phi\,\nabla\!\nabla \phi;\;\;\;\;\;\;\;\;\;\;\;\;
|\dot \xb|^2 - |\nabla \phi|^2 = H,\label{eq:path}
\end{equation} where $\nabla\!\nabla \phi$ is the Hessian matrix of $\phi$.
These correspond to the (conservative) Hamiltonian motion of a particle of mass 2 (actually 2$\times$ friction${}^2$) moving in an effective potential $F=- |\nabla \phi|^2$, and the quantity $H$ is conserved along the path \cite{ge}.
$H$ is analogous to the energy in the effective system, but has the dimensions of a power. Note that here we are really saying that the dominant (non-differentiable) paths lie within a small tube around the solution to (\ref{eq:path}) \cite{stratonovich} and fluctuations around this can be integrated over to determine the pre-exponential (entropic) factor in the transition rate -- see e.g.\ \cite{schulman}.
Here we focus on determining the MLPs, rather than the rates themselves.
Inserting Eq.~\eqref{eq:path} into the action integral yields
\begin{equation}
S  =  2\left[\phi(\xb_1)-\phi(\xb_0)\right] - Ht + 2\int_{\gamma} \sqrt{H+|\nabla \phi|^2}|\dee\xb|,
\end{equation}
with the time for the path given by
\begin{equation}
t  =  \int_{\gamma} \frac{|\dee\xb|}{\sqrt{H+|\nabla \phi|^2}}.
\label{eq:time}\end{equation} $S$ is Hamilton's principal function for the effective classical mechanics, and corresponds to the large deviations rate function for the stochastic dynamics.
$\gamma$ is the optimal path through the potential linking $\xb_0$ and $\xb_1$, i.e.\ is the solution of (\ref{eq:path}).
The relation between the path power $H$ and the time $t$ comes from either solving the classical equation of motion, or extremizing $S$ over $H$.
$t\to\infty$ corresponds to $H\to 0$, provided the path includes a critical point of $\phi$, which is the case for the transitions of interest.
When $H=0$, $\dot \xb=\pm\nabla \phi$, and $\gamma$ is the minimum energy path, which can be determined using e.g.\ gMAM \cite{vanden2008geometric}.
This path corresponds to $t\to\infty$ and the long-time average rate, since 
\begin{equation}
S \to 2\Delta\phi + 2 \int_{\gamma}|\nabla \phi|\,|\dee \xb |\;\; = \;\; \begin{cases}
  0 & \text{downhill path}\\
            4\Delta\phi>0 & \text{uphill path.}
  \end{cases}
\end{equation}
The last equality follows from the fact that, for $H=0$, the path is always (anti-)parallel to $\nabla \phi$ (note that the converse to this statement is not necessarily true).
This zero-power path, the instanton, recovers the familiar Kramers form $\exp(-\beta\Delta\phi)$ for the average rate at which an energy barrier of height $\Delta\phi$ is traversed.
Different values of $H$ correspond to different paths -- the equation of motion $\ddot \xb = -\nabla F$ has different boundary conditions.
We now refer to the path as $\gamma_H$, and note that $\gamma_0$ is the absolute minimum action path determined by the original gMAM algorithm.
In particular, the initial and final velocity vectors for $\gamma_H$ have different magnitudes and directions from those of $\gamma_0$, which start and end at rest. 

The gMAM algorithm \cite{vanden2008geometric,heymann2008geometric,leines2016comparison} can be modified to include paths with nonzero power $H$ as follows. Following the notation of \cite{vanden2008geometric}, parameterize the curve $\gamma_H$ by normalized arc length using $\alpha\in [0,1]$, and let $\XX(\alpha) = (X_1(\alpha),X_2(\alpha),...)$ be the parametric equations of the curve. The path-dependent part of the action can be written as 
\eqa
S-2\Delta\phi & = & -Ht + 2\int_{\gamma} \sqrt{H+|\nabla \phi|^2}|\dee\XX|\nonumber\\ 
& = & \int_{\gamma} \frac{H+2|\nabla \phi|^2}{\sqrt{H+|\nabla \phi|^2}}|\dee\XX|\nonumber\\ 
& = & \int_0^1g(\alpha)\,\XX'^2(\alpha)\,\dee\alpha, 
\eeqa
where the prime denotes differentiation with respect to $\alpha$, and 
\eq
g(\alpha) = \frac1{|\XX'(\alpha)|}\frac{H+2|\nabla \phi|^2}{\sqrt{H+|\nabla \phi|^2}}. 
\eeq 
The Euler-Lagrange equation for $\XX(\alpha)$ then reads
\begin{eqnarray}
\frac{\delta S}{\delta X_i}
= \frac1{g}\left(\nabla \phi\right)_j \left(\nabla\nabla \phi\right)_{ji}\!& &\!\frac{(H+2|\nabla \phi|^2)(3H+|\nabla \phi|^2)}{(H+|\nabla \phi|^2)^2} \notag\\
& & - \left(gX'_i\right)' \;\;\;\;=\;\;\;\; 0,  
\end{eqnarray} and $\XX$ is evolved from an initial guess (e.g.\ the straight line from $\xb_0$ to $\xb_1$) according to
\eq
\frac{\dee\XX}{\dee\tau} = -g\frac{\delta S}{\delta\XX}.
\eeq
The factor of $g>0$ avoids potential numerical issues when $H$ and $\nabla \phi$ become small. Full details can be found in \cite{vanden2008geometric}, where the authors also present a robust and efficient numerical implementation that avoids the computation of the Hessian $\nabla\!\nabla \phi$.
Note Ref.~\cite{leines2016comparison} investigates the convergence of gMAM as compared with the string method, finding that gMAM more reliably identifies the MEP in complex landscapes. 

Because $H$ is defined implicitly in Eq.~(\ref{eq:time}), and the path $\gamma_H$ depends on $H$, it cannot be determined \emph{a priori}. If the time $t$ is specified, a further iterative process is required to find $H$. For simple 1D paths, $t$ is a decreasing function of $H$, but in higher dimensions it is not as simple, since different values of $H$ can produce very different paths. 
As $H$ becomes large, it is much greater than all values of $|\nabla \phi|^2$, and so the path $\gamma_H$ becomes the straight line from $\xb_0$ to $\xb_1$.

In Fig.~\ref{fig:2D}(a--b) we display the MLP for various $H$, i.e.\ for various values of $t$, obtained from our extended gMAM algorithm.
The blue $H=0$ path corresponds to $t\to\infty$, which is the MLP predicted by RRT.
As $H$ is increased, we see from Fig.~\ref{fig:2D}(a) that the MLP no longer passes through the saddle point (the transition state of RRT) at $\xb_D=(-1,0)$, instead cutting the corner.
For $H\geq205$ we see that the MLP jumps to the other side of the potential and no longer goes anywhere near point D and instead goes in the vicinity of point E, i.e.\ over the much higher energy barrier; see e.g.\ the $H=205$ red path in Fig.~\ref{fig:2D}(a), which has corresponding time $t=0.047$. For times of order $t\sim0.1$ we find that paths via either route have roughly the same path action, despite having very different barrier heights.
The order of magnitude of this time is in agreement with what we see in Fig.~\ref{fig:2D}(c), from solving Eq.~\eqref{eq:FP}, i.e.\ the time when the densities at the two saddle points are equal.
In the SI we give the values of $H$ used together with the corresponding times $t$. 

We have shown that the important difference between finite-time minimum action paths and their infinite-time limit, the instanton, is that conditioned on a finite time, the minimum action (and hence most probable) path need not traverse the lowest energy barrier.
Although the instanton is the path from $\xb_0$ to $\xb_1$ involving the absolute minimum of hill-climbing, when constrained to a finite time, a shorter path may be worth the extra uphill. 
This has implications for any stochastic transition where only a finite time is available for the reaction to occur, particularly if there are several paths the system can take.
Moreover, transition paths and energy barriers inferred from experiments or simulations conducted over too short a time scale could easily be very different from the paths and barriers that dominate the system dynamics in reality.

\acknowledgments

SPF acknowledges support from the UK EPSRC, grant number EP/R005974/1. We benefited from valuable discussions with Celia Reina, Tapio Ala-Nissila, Thomas Bartsch, Rob Jack and Uwe Thiele.


%


\clearpage

\onecolumngrid

\section*{Supplementary information for:\\
    Stochastic transitions: Paths over higher energy barriers can dominate in the early stages}

\renewcommand{\thefigure}{A\arabic{figure}}
\setcounter{figure}{0}
    
\begin{center}
S.~P. Fitzgerald, A. Bailey Hass, G. D{\'\i}az Leines and A.~J. Archer
\end{center}

\vspace{1cm}


\section{1D Model potential}
The 1D potential that we consider (displayed in Fig.~1 of the main text) is:
\begin{equation}
\beta\phi(x)=\left(\frac{x}{3}\right)^{10}+e^{-2(x+1)^2}+3e^{-12(x-2)^2}-\frac{13}{10}e^{-12(x-2.5)^2}-\frac{3}{10}e^{-8(x-1)^2}-\frac{23}{10}e^{-8(x+2)^2}.
\nonumber
\end{equation}
Precise locations of the 3 minima are:\\
$x=x_B=-2.0117264,$ (the global minimum)\\
$x=x_A=1.0004288,$ (a local minimum and our start point)\\
$x=x_C=2.5229797$ (a local minimum)\\
and the two local maxima are at:\\
$x=x_D=-0.99708709$\\
$x=x_E=1.9912978.$

\section{2D Model potential}

The 2D potential that we consider (displayed in Fig.~2 of the main text and in Fig.~\ref{fig:A1} below) is:
\begin{eqnarray}
\beta\phi(x,y)=4(x^2+4y^2-1)^2-\frac{1}{2}x-2e^{-4(x-\frac{1}{2})^2-4(y-\frac{1}{2})^2}
-e^{-4(x-\frac{1}{2})^2-4(y+\frac{1}{2})^2}+3e^{-4(x-1)^2-4y^2}.
\nonumber
\end{eqnarray}

\begin{figure}[b!]
\centering
\includegraphics[width=0.6\textwidth]{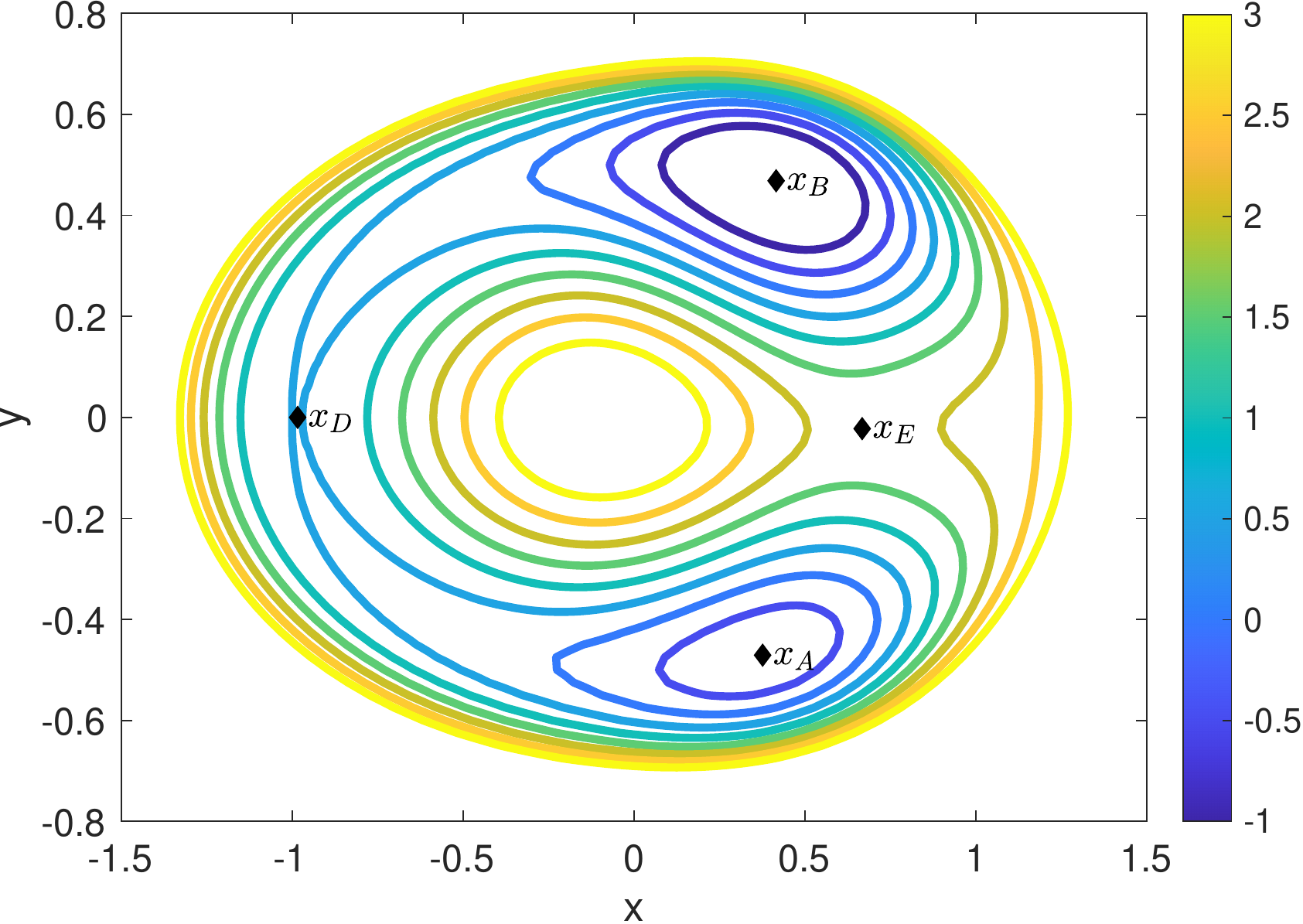}
\caption{The 2D model potential}
\label{fig:A1}
\end{figure}

This potential has two minima at:\\
$\xb_A=(x_A,y_A)=(0.37610659,-0.47094557)$, (a local minimum and our start point)\\
$\xb_B=(x_B,y_B)=(0.41564013,0.46836780)$, (the global minimum).

The 2D potential has a local maximum near the origin at:\\
$\xb=(-0.097396665,-0.0053727757)$\\
and there are two saddle points at:\\
$\xb_D=(x_D,y_D)=(-0.98392863,-0.00010774153)$\\
$\xb_E=(x_E,y_E)=(0.66721235,-0.022398035)$.

\vspace{1cm}
In Table~\ref{table:HtSv} below we give the times $t$ corresponding to various values of the path power $H$. Some of these paths are displayed in Fig.~2 of the main text.

\begin{table}[h!]
\centering
\begin{tabular}{|c|c|c|c|c|c|c|c|c|c|c|c|c|c|c|c|c|c|c|c|c|c|c|}
\hline
     $H$ & 0 & 0.01 & 0.02 & 0.03 & 0.1 & 2.5 & 5.0 & 10.0 & 25.0 & 50.0 & 100.0 & 195.0 & 200.0 & 205.0 & 212.0 & 235.0 & 245.0 & 250.0 & 300.0 & 400.0 & 500.0 & 1000.0 \\
     \hline
     $t$  & $\infty$ & 7.84 & 6.85 & 6.28 & 4.70 & 1.61 & 1.21 & 0.888 & 0.575 & 0.405 & 0.277 & 0.153 & 0.165 & 0.0478 & 0.0463 & 0.0419 & 0.0402 & 0.0413 & 0.0338 & 0.0255 & 0.0221 & 0.0168 \\
     \hline
\end{tabular}
\caption{Time $t$ for various $H$ values.}
\label{table:HtSv}
\end{table}

\begin{figure}
\centering
\includegraphics[width=0.5\textwidth]{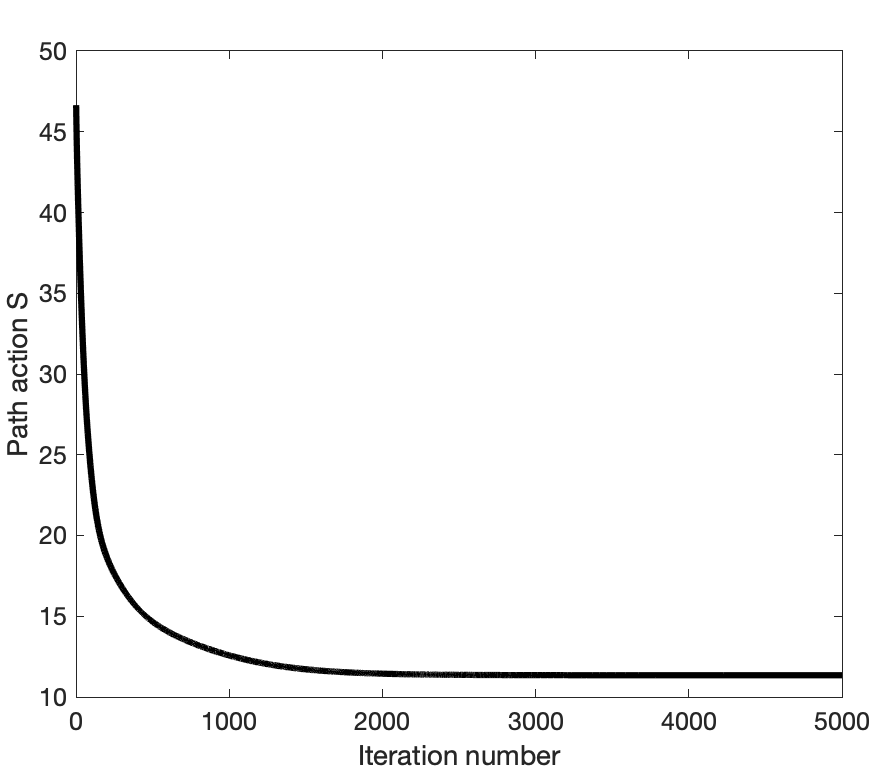}
\caption{Convergence of the extended gMAM algorithm to find the MLP for $H=0$.}
\label{fig:conv}
\end{figure}

Fig.~\ref{fig:conv} shows the algorithm converging to the minimum action path (the MLP) for $H=0$ from the straight line initial guess.

\end{document}